\documentclass[aps,prb,twocolumn,nofootinbib,superscriptaddress]{revtex4-2}
\usepackage[]{graphicx,hyperref}
\usepackage{dcolumn}
\usepackage{bm}
\usepackage{amsmath}
\usepackage{braket}
\usepackage{physics}
\usepackage{amsfonts}
\usepackage{amssymb}
\usepackage{subfigure}
\usepackage{multirow}
\newcommand{\nn}{\nonumber}

\hypersetup{
setpagesize=false,
 bookmarksnumbered=true,
 bookmarksopen=true,
 colorlinks=true,
 linkcolor=black,
 citecolor=blue,
}

\begin{document}
\preprint{APS/123-QED}
\title{Pairing symmetries of multiple superconducting phases in UTe$_2$:\\
        Competition between ferromagnetic and antiferromagnetic fluctuations}
\author{Jushin Tei}
\email{tei@blade.mp.es.osaka-u.ac.jp}
\affiliation{Department of Materials Engineering Science, Osaka University, Toyonaka 560-8531, Japan}
\author{Takeshi Mizushima}
\affiliation{Department of Materials Engineering Science, Osaka University, Toyonaka 560-8531, Japan}
\author{Satoshi Fujimoto}
\affiliation{Department of Materials Engineering Science, Osaka University, Toyonaka 560-8531, Japan}
\date{\today}

\begin{abstract}
The putative spin-triplet superconductor UTe$_2$ exhibits multiple superconducting phases under applied pressure [D. Braithwaite {\it et al}., Commun. Phys. {\bf 2}, 147 (2019)]. 
The clarification of pairing mechanisms and symmetries of gap functions are essentially important for understanding the multiple-phase diagram.
Since the coexistence of ferromagnetic and antiferromagnetic spin fluctuations with Ising-like anisotropy is suggested from measurements of magnetic susceptibilities and
neutron scattering measurements, it is expected that
the interplay between these spin fluctuations plays a crucial role in the emergence of 
the multiple superconducting phases.
Motivated by these observations,  
we examine the spin-fluctuation-mediated pairing mechanism, analyzing the linearized Eliashberg equations for an effective model of $f$-electron bands.
It is found that the Ising-like 
ferromagnetic fluctuations stabilize spin-triplet pairings in either the $A_u$ or $B_{3u}$ states, 
whereas Ising-like antiferromagnetic fluctuations stabilize spin-triplet pairings in the $B_{1u}$ state.
These results provide a plausible scenario elucidating the multiple superconducting phases under pressure.

\end{abstract}

\maketitle

\section{Introduction}
The heavy fermion system UTe$_2$ has been gathering significant attention 
due to its distinctive superconducting properties~\cite{ran2019Nearly,aoki2022Unconventional}.
Notably, it exhibits an upper critical field that far exceeds the Pauli limit~\cite{ran2019Nearly,aoki2019Unconventional,knebel2019FieldReentrant}, 
reentrant and reinforcement behaviors in response to magnetic fields~\cite{knebel2019FieldReentrant,ran2019Extreme,rosuel2023FieldInduceda,wu2023Enhanced},
and multiple superconducting phases under both magnetic fields and pressure~\cite{ran2019Extreme,rosuel2023FieldInduceda,braithwaite2019Multiple,aoki2020Multiple,thomas2020Evidence,knebel2020Anisotropy,lin2020Tuning,ran2020Enhancement,aoki2021FieldInduced}.
These remarkable features strongly suggest the potential realization of a spin-triplet pairing state,
holding significant promise as a topological superconductor.
The existence of Majorana surface states in UTe$_2$ has been predicted in previous studies~\cite{ishizuka2019InsulatorMetal,tei2023Possible}.

However, the determination of the symmetries of the gap functions, which is crucial in defining the superconducting properties,
remains a subject of controversy.
In earlier studies, spontaneous time-reversal symmetry breaking (TRSB) has been discussed from scanning tunneling microscopy (STM) measurements~\cite{jiao2020Chiral} and Kerr-effect measurements~\cite{hayes2021Multicomponent}.
However, in contrast, recent Kerr-effects measurements utilizing high-quality samples have shown no evidence for a spontaneous Kerr signal~\cite{ajeesh2023Fatea}, leaving no conclusive empirical support for TRSB.

In the case of spin-triplet pairings expected for UTe$_2$, 
possible symmetries are the \{$A_u$, $B_{1u}$, $B_{2u}$, $B_{3u}$\} states, which are irreducible representations of the point group $D_{2h}$.
Extensive experimental studies have been performed to elucidate the gap symmetry.
The specific heat measurements~\cite{rosa2022Single, ishihara2023Chiral} and magnetic penetration depth measurements~\cite{ishihara2023Chiral} 
provide support for the existence of point nodes, indicating the $B_{3u}$ state.
In contrast, thermal conductivity measurements support a full gap state~\cite{suetsugu2024Fully}.
Also, the nuclear magnetic resonance (NMR) measurements~\cite{matsumura2023Large} demonstrate a reduction in Knight-shift for all directions, suggesting the $A_u$ state without nodes.
The notable recent improvement in sample quality~\cite{sakai2022Single} holds promising prospects for further progress in experimental research.

\begin{figure}[tb]
    \centering
    \includegraphics[width=\linewidth]{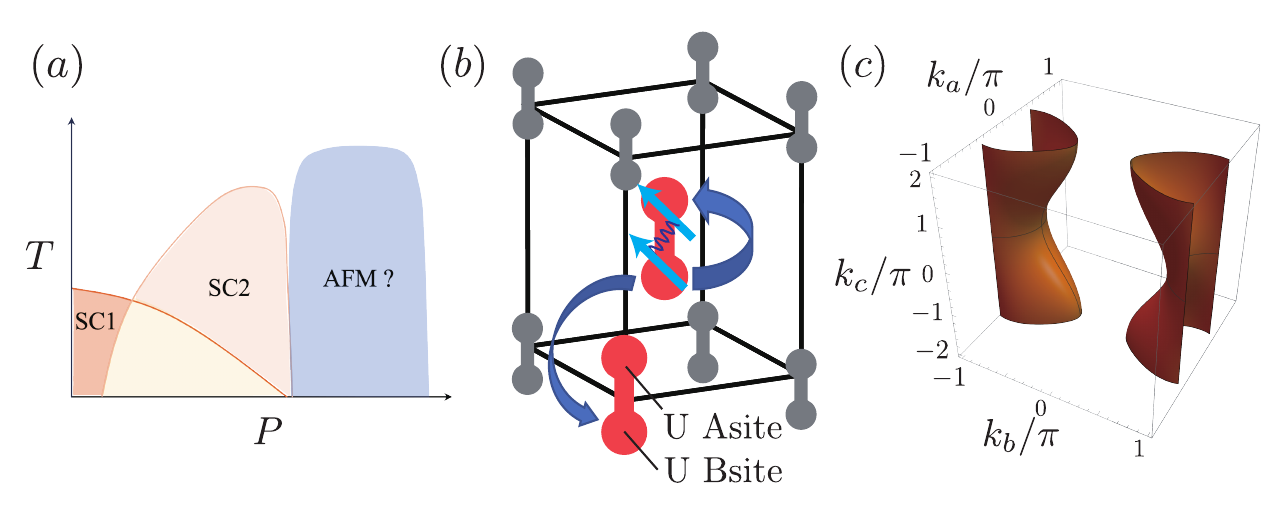}
    \caption{(a) Schematic $P$-$T$ superconducting phase diagram.
    (b) Crystal structure of UTe$_2$. Only U sites are depicted.
    (c) Cylindrical electron Fermi surface of the two-orbital model of UTe$_2$.}
    \label{fig:model}
\end{figure}

Another striking feature of UTe$_2$ is the occurrence of multiple superconducting phases under applied pressure~\cite{braithwaite2019Multiple,aoki2020Multiple}.
The schematic pressure-temperature phase diagram is depicted in Fig~\ref{fig:model}(a).
The superconducting phase at ambient pressure denoted as SC1 is suppressed as applied pressure increases,
whereas another superconducting phase, labeled as SC2, appears under pressure and abruptly vanishes at around 1.5 GPa, coinciding with the appearance of a putative antiferromagnetic ordered phase~\cite{aoki2020Multiple,thomas2020Evidence,li2021Magnetica,valiska2021Magnetic}.

To shed light on the nature of the multiple superconducting phases, a comprehensive understanding of the pairing origin is essential.
In earlier studies~\cite{ran2019Nearly}, an anisotropic increase in uniform magnetic susceptibility was observed, 
signifying the presence of Ising-like ferromagnetic (FM) fluctuations 
and indicating their potential as a pairing glue.
However, direct observation of the FM fluctuations has not been achieved hitherto.
In contrast, neutron scattering measurements have revealed the presence of antiferromagnetic (AFM) fluctuations,
characterized by the ordering vector $Q \approx (0,\pi,0)$~\cite{duan2020Incommensuratea,knafo2021Lowdimensional,raymond2021Feedback}.
These experimental observations strongly suggest a coexistence of FM and AFM fluctuations,
which may hold the key to unraveling the intricate nature of the multiple superconducting phases under pressure.

Motivated by these considerations, we postulate that pairing glues are both FM and AFM fluctuations, and
analyze the linearized Eliashberg equations to determine stable pairing states under pressure.
We introduce a pressure parameter that controls the interplay between FM and AFM fluctuations, 
thereby reproducing the multiple superconducting phases under applied pressure.

The organization of this paper is as follows.
In Sec.~II we briefly describe a theoretical model of superconductivity in UTe$_2$ and a theoretical calculation method.
In Sec.~III we present the calculated results obtained by analyzing the linearized Eliashberg equations, and discuss the relationship between characters of spin fluctuations and stable pairing symmetries. 
Based on numerical results, we provide a scenario for understanding the multiple superconducting phases
under pressure.
A comprehensive discussion and summary are presented in Sec. IV.

\section{Model and Calculation Method}
To elucidate the superconducting states of UTe$_2$,
we construct a minimal model that integrates crucial properties of UTe$_2$,
including the orbital degrees of freedom, the Fermi surface observed in de Haas-van Alphen (dHvA) experiments~\cite{aoki2022First,eaton2024Quasi2D}, magnetic anisotropy, and magnetic spin fluctuations.
In this section, we explain the employed model and the basic formulations for the calculations of stable superconducting states.

\subsection{Minimal model of superconductivity in UTe$_2$}
Here, we explain the tight-binding model utilized for investigating superconductivity in UTe$_2$.
UTe$_2$ has a body-centered orthorhombic lattice structure, characterized by the space-group $Immm (\#71, D^{25}_{2h})$.
Within the unit cell, two U atoms organize dimers arranged along the $c$-axis,
and these dimer chains extend parallel to the $a$-axis, as depicted in Fig.~\ref{fig:model}(b).
The first principle calculations suggest that $f$-orbital electrons in the dimer chains are crucial for the band structure~\cite{xu2019QuasiTwoDimensionala,shishidou2021Topological}.
Motivated by these observations, we consider a two-orbital system as a minimal model.
Following Ref~\cite{shishidou2021Topological}, we exploit the non-interacting Hamiltonian for the normal state with all symmetry-allowed terms,
\begin{eqnarray}
    \label{eq:Hn}
    H_N(\bm{k}) = (\epsilon_0(\bm{k}) - \mu)\sigma_0\tau_0 + f_x(\bm{k})\sigma_0\tau_x  \nn \\
    + f_y(\bm{k})\sigma_0\tau_y + \bm{g(k)}\cdot\boldsymbol{\sigma}\tau_z,
\end{eqnarray}
with,
\begin{eqnarray}
    &&\epsilon_0(\bm{k}) = 2t_1\cos k_a + 2t_2\cos k_b, \\
    &&f_x(\bm{k}) = t_3 + t_4 \cos(k_a/2)\cos(k_b/2)\cos(k_c/2), \\
    &&f_y(\bm{k}) = t_5 \cos(k_a/2)\cos(k_b/2)\sin(k_c/2), \\
    &&g_a(\bm{k}) = R_a \sin k_b, \\
    &&g_b(\bm{k}) = R_b \sin k_a, \\
    &&g_c(\bm{k}) = R_c \sin(k_a/2)\sin(k_b/2)\sin(k_c/2), 
\end{eqnarray}
where $\sigma_{a,b,c}$ and $\tau_{x,y,z}$ are Pauli matrices for spin and orbital degrees of freedom, respectively, and
 $\sigma_0$ and $\tau_0$ are $2 \times 2$ unit matrices.
In this paper, we use the labels $a, b$, and $c$ to represent crystalline axes, and the labels $x, y$, and $z$ are employed to represent orbital degrees of freedom.
The last term in Eq.~(\ref{eq:Hn}) is the staggered Rashba spin-orbit interaction, arising from the local inversion symmetry breaking.
To reproduce the cylindrical electron Fermi surface illustrated in Fig.~\ref{fig:model}(c), which is suggested by the dHvA experiments~\cite{aoki2022First},
we choose the band parameters as follows: 
$\mu=-3.6$, $t_1 = -1.0$, $t_2=0.75$, $t_3 = -1.4$, $t_4 = 1.3$, $t_5 = -1.3$, $R_a = R_b = R_c = 0.2$.
The presence of a cylindrical hole Fermi surface has also been reported in the dHvA experiments.
However, according to the first principle band calculations~\cite{ishizuka2021Periodic,xu2019QuasiTwoDimensionala}, the hole band primarily consists of $d$-electrons of U sites and $p$-electrons of Te sites.
Considering the substantive role played by strongly correlated $f$-electrons in magnetic fluctuations, 
we expect that superconductivity is also mainly triggered by $f$-electrons, and 
disregards the influence from the hole band, opting for a two-orbital model involving $f$-electrons.
This minimal model is sufficient for our purpose of determining the most stable pairing states induced by FM and AFM spin fluctuations arising from
$f$-electrons.

\begin{table}[t]
    \centering
    \caption{Irreducible representations (IRs) of the point group $D_{2h}$ and basis functions.
    In the two-orbital system, the gap functions take the form
    $\tau_j\otimes\{i(\bm{d}_j\cdot\boldsymbol{\sigma} + \psi_j)\sigma_b\}$,
    where $\tau_j~(j=0,x,y,z)$ are Pauli matrices for orbital degrees of freedom.
    $\bm{d}$ ($\psi$) represents the spin-triplet (spin-singlet) component.}
    \begin{tabular}{cccc}
    \hline\hline
      & \multicolumn{3}{c}{Basis function} \\ \cline{2-4}
      ~~~IRs~~~   & \multicolumn{3}{c}{Orbital degrees of freedom} \\
        & $\tau_0,\tau_x$        & $\tau_y$        & $\tau_z$        \\ \hline
    $A_u$  &  $\bm{d} = k_a\hat{a}+k_b\hat{b}+k_c\hat{c}$      &   $\bm{d} = \hat{c}$         &   $\psi =  k_ak_b$        \\
    $B_{1u}$ & $\bm{d} = k_b\hat{a}+k_a\hat{b}$         &            &    $\psi =  1$        \\
    $B_{2u}$ &  $\bm{d} = k_c\hat{a}+k_a\hat{c}$        &   $\bm{d} = \hat{c}$         &    $\psi =  k_bk_c$        \\
    $B_{3u}$ & $\bm{d} = k_c\hat{b}+k_b\hat{c}$         &   $\bm{d} = \hat{c}$         &    $\psi =  k_ck_a$        \\ \hline
    $A_g$  &   $\psi =  1$       &   $\psi=k_c$         &    $\bm{d} = k_b\hat{a}+k_a\hat{b}$        \\
    $B_{1g}$ & $\psi =  k_ak_b$        &        &   $\bm{d} = k_a\hat{a}+k_b\hat{b}+k_c\hat{c}$         \\
    $B_{2g}$ &  $\psi =  k_ck_a$        &  $\psi=k_a$          &   $\bm{d} = k_c\hat{b}+k_b\hat{c}$         \\
    $B_{3g}$ &  $\psi =  k_bk_c$        &  $\psi=k_b$          &   $\bm{d} = k_c\hat{a}+k_a\hat{c}$         \\ \hline\hline
    \end{tabular}
    \label{IRs}
\end{table}

The pairing states of UTe$_2$ are classified by the point group $D_{2h}$,
encompassing eight irreducible representations:
$A_g$, $B_{1g}$, $B_{2g}$, $B_{3g}$, $A_u$, $B_{1u}$, $B_{2u}$, and $B_{3u}$.
In the two-orbital model, the gap functions also have orbital degrees of freedom, 
and take the following form, 
\begin{eqnarray}
    \Delta = \sum_{j=0,x,y,z} \tau_j \otimes \{ (\bm{d}_j\cdot\boldsymbol{\sigma} + \psi_j)i\sigma_b \},
\end{eqnarray}  
where $\bm{d} = (d^a,d^b,d^c)$ represents a $d$-vector for spin-triplet pairings and $\psi$ is a component of a spin-singlet pairing state.
In the following, we refer to $(d^0\equiv \psi, d^a, d^b, d^c)$ as $d$-vector components.
The main basis functions of $d$-vector components for each irreducible representation are summarized in Table~\ref{IRs}.
Note that the mixing of spin-triplet pairings $\bm{d}$ and spin-singlet pairings $\psi$ occurs in intraorbital pairing channels due to local inversion breaking at each U atomic site.
Since inversion symmetry operation exchanges the two orbitals, i.e. A and B sites in the unit cell, as seen in Fig~\ref{fig:model}(b), 
global inversion symmetry implies, 
\begin{eqnarray}
    \{\Delta (\bm{k})\}_{\rm AA} = \pm \{\Delta (-\bm{k})\}_{\rm BB}, \\
    \{\Delta  (\bm{k})\}_{\rm AB}= \pm \{\Delta (-\bm{k})\}_{\rm BA},
\end{eqnarray}
where $+$ ($-$) sign is for the $g(u)$ irreducible representations.

\subsection{Pairing interaction}
\begin{table}[t]
    \caption{Form of the effective interaction which couples to each $d$-vector component.}
    \label{table:vertex}
    \centering
    \begin{tabular}{cc}
    \hline
    \hline
    \hspace{15pt}$d$-vector component\hspace{15pt} & \hspace{15pt}Effective interaction\hspace{15pt} \\ \hline
    $d^0(\psi)$ & $-\chi^a-\chi^b-\chi^c$              \\
    $d^a$ & $-\chi^a+\chi^b+\chi^c$              \\
    $d^b$ & $+\chi^a-\chi^b+\chi^c$              \\
    $d^c$ & $+\chi^a+\chi^b-\chi^c$              \\ \hline\hline
    \end{tabular}
\end{table}
As mentioned in the Introduction, 
the superconductivity of UTe$_2$ occurs close to  both ferromagnetic and antiferromagnetic critical points.
Additionally, applied pressure induces the magnetically ordered phase.
From these perspectives, it is plausible to consider that the competition between FM and AFM fluctuations, controlled by pressure,
gives rise to the multiple superconducting phases under pressure.
Guided by this consideration, we postulate that pairing glues are FM and AFM fluctuations
and examine pairing states induced by these spin fluctuations.
To delineate these interactions, we employ an effective Hamiltonian 
with the following form of spin-spin interactions, 
\begin{eqnarray}
    \label{eq:int}
    \hat{V} = -\sum_{\mu, q}\chi^\mu(q)\hat{S}^\mu(q)\hat{S}^\mu(-q).
\end{eqnarray}
Here, $q = (\bm{q},i\Omega_n)$, and $\Omega_n = 2n\pi T$ represents the bosonic Matsubara frequency.
$\hat{S}^\mu$ $(\mu= a,b,c)$ is the spin operator expressed as
\begin{eqnarray}
    \label{eq:spin}
    \hat{S}^{\mu}(q) = \sum_{k,\lambda,\alpha,\beta} c^\dagger_{k+q,\lambda\alpha}\sigma^{\mu}_{\alpha\beta}c_{k,\lambda\beta},
\end{eqnarray}
where $k=(\bm{k},i\omega_n)$, $\omega_n = (2n+1)\pi T$ represents the fermionic Matsubara frequency,
and $c_{k,\lambda\alpha}$ ($c^\dagger_{k,\lambda\alpha}$) is an annihilation (creation) operator of an $f$-electron with momentum $\bm{k}$ and Matsubara frequency $\omega_n$, and spin $\alpha$ on an atomic site $\lambda = \rm A, B$.
$\chi^\mu(q)$ is the magnetic susceptibility, as given by the phenomenological form following Ref.~\cite{monthoux1999Wave}:
\begin{eqnarray}
    \label{eq:chi}
    \chi^\mu(q) = \frac{J^\mu}{1/{\xi^\mu}^2 + (\hat{\bm{q}}-\hat{\bm{Q}})^2 + |\Omega_n|\hat{q}^{2-z}/T_{\rm{sf}}},
\end{eqnarray}
where $J^\mu = g^2\chi_0/(\xi^{\mu}_0)^2$, $g$ is the coupling constant of interacting strength,
$\xi^\mu$ ($\xi^{\mu}_0$) is the correlation length with (without) strongly correlated effects,
$\hat{\bm{Q}}$ is the ordering vector defined as $\bm{Q} = \bm{0}$ ($\bm{Q} = (0,\pi,0)$) for FM (AFM) fluctuations, 
$z$ is the dynamical exponent $z=3$ ($z=2$) for the FM (AFM) fluctuations, and 
$T_{\rm{sf}}$ is a characteristic spin fluctuation temperature and we set $T_{\rm{sf}}=4$.
For Eq.~(\ref{eq:chi}) to satisfy body-centered periodicity,
let the momentum dependence for FM fluctuations be replaced as,
\begin{eqnarray}
    (\hat{\bm{q}}-\hat{\bm{Q}})^2  &\Rightarrow& 8 - 8\cos\frac{q_a}{2}\cos\frac{q_b}{2}\cos\frac{q_c}{2}, 
\end{eqnarray}
and for AFM fluctuations be replaced as,
\begin{eqnarray}
   ( \hat{\bm{q}}-\hat{\bm{Q}})^2  &\Rightarrow& 8 - 8\cos\frac{q_a}{2}\cos\frac{q_b\pm\pi}{2}\cos\frac{q_c}{2}.
\end{eqnarray}
Note that, in the above notation, length is nondimensionalized in units of lattice constants $L_\mu$, thus $\hat{\bm{q}}$ is also a dimensionless quantity whose unit is $L_\mu^{-1}$.
We assume $J^a \gg J^b,~J^c$ due to the considerable Ising anisotropy observed in UTe$_2$~\cite{ran2019Nearly}.
For simplicity, we assume that the magnetic susceptibility is independent of the atomic sites of U denoted by $\lambda$.
\color{black}

\subsection{Eliashberg equation}
To discuss instability toward superconducting states, we analyze the linearized Eliashberg equations:
\begin{eqnarray}
    \label{eq:Eliash}
    \varepsilon\Delta_{\zeta\zeta'}(k) = - \frac{T}{N}&&\sum_{k',\zeta_1,\zeta_2,\zeta_3,\zeta_4}V_{\zeta\zeta_1;\zeta'\zeta_2}(k-k') \nn \\
    &&\times G_{\zeta_1\zeta_3}(k')\Delta_{\zeta_3\zeta_4}(k')G_{\zeta_2\zeta_4}(-k'),
\end{eqnarray}
\begin{eqnarray}
    \label{eq:Green}
    G(k) = [i\omega_n - H_N(\bm{k})-\Sigma(k)]^{-1},
\end{eqnarray}
\begin{eqnarray}
    \label{eq:self}
    \Sigma_{\zeta\zeta'}(k) = -\frac{T}{N}\sum_{k',\zeta_1,\zeta_2}V_{\zeta\zeta_1;\zeta'\zeta_2}(k-k')G_{\zeta_1\zeta_2}(k'),
\end{eqnarray}
where label $\zeta = (\lambda,\alpha)$ denotes orbital and spin degrees of freedom,
$G_{\zeta\zeta'}(k)$ is the single-particle Green's function,
and $\Sigma_{\zeta\zeta'}(k)$ is the quasiparticle self-energy. 
The explicit expression of $G_{\zeta\zeta'}(k)$ is presented in Appendix A.
Eq.~(\ref{eq:Eliash}) determines the transition temperature, when the eigenvalue condition $\varepsilon(T) \rightarrow 1$ is satisfied.
However, to reduce computational cost,
we keep the temperature at a constant value of $T = 0.02|t_1|$ for solving the linearized Eliashberg equations,
and compare the maximum eigenvalues $\varepsilon_{\rm{max}}$ of all the irreducible representations to determine the most stable pairing state.

Because of spin-orbit interactions, Eq.~(\ref{eq:Eliash}) is a set of coupled equations encompassing all $d$-vector components.
Each  $d$-vector component is coupled to a specific form of the effective interaction $V_{\zeta\zeta_1;\zeta'\zeta_2}(q)$, as summarized in Table~\ref{table:vertex}.

\section{Numerical Results of Stable Pairing States}
\begin{figure}[t]
    \centering
    \includegraphics[width=\linewidth]{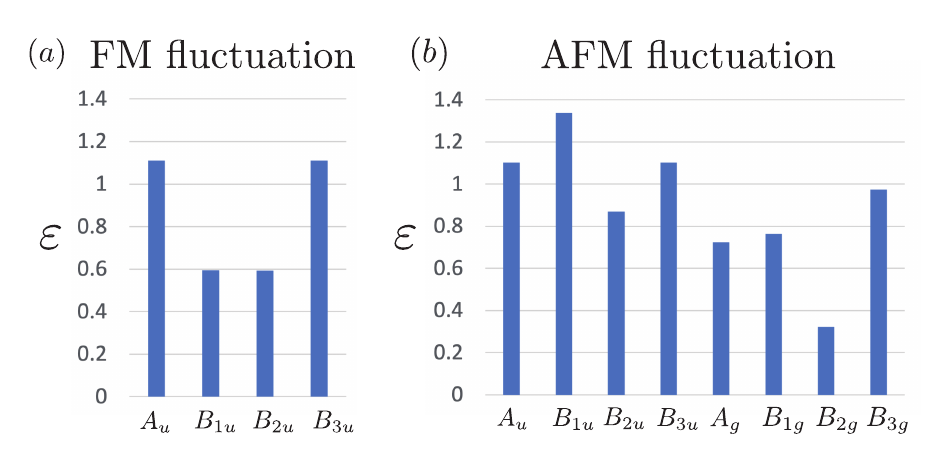}
    \caption{The largest eigenvalue of the Eliashberg equation for each irreducible representation in the case of Ising (a)~FM and (b)~AFM fluctuations, respectively.
    The interaction parameters are set as, (a)~$J^a_{\text{FM}} = 30$~(b)~$J^a_{\text{AFM}} = 100$,~$J^b_{\text{FM/AFM}} = 0$,~$J^c_{\text{FM/AFM}} = 0$, and $(1/{\xi^a})^2 = 0.1$.
    In the case of (a)~FM fluctuations, the eigenvalues of the $A_u$ and $B_{3u}$ states are the largest values and nearly degenerate.
    All the even-parity representations do not have finite eigenvalues.
    In the case of (b)~AFM fluctuations, the eigenvalue of the $B_{1u}$ state is the largest.
     }
    \label{fig:FMAFM}
\end{figure}

\begin{table}[t]
    \centering
    \begin{minipage}[t]{0.45\textwidth}
    \centering
    \caption{Norms of $d$-vector components $|d^\alpha| = \sqrt{\sum_{\bm{k},n,j}|d^\alpha_j(\bm{k},i\omega_n)|^2}$  which are defined by the sum of all orbital contributions\color{black}, in the case of FM fluctuations.
    Bold types represent predominant components.
    Ising FM fluctuations orient the $d$-vector toward the $b$- or $c$-axis.
    Namely, the spin momentum of Cooper pairs aligns along the $a$-axis.}
    \label{table:FM}
    \begin{tabular}{ccccc}
        \hline\hline
            & $|\psi|$ & $|d^a|$ & $|d^b|$ & $|d^c|$ \\ \hline
        $A_u$  & 0.04 & 0.01   & $\bm{0.99}$  & 0.05      \\
        ~$B_{1u}$~ & ~0.10~ & ~0.12~ & ~{0.13}~   &  ~$\bm{0.98}$~    \\
        $B_{2u}$ & 0.13 & 0.11 & $\bm{0.98}$   &  {0.10}    \\
        $B_{3u}$ & 0.01 & 0.04   & 0.05   &  $\bm{0.99}$     \\ \hline\hline
    \end{tabular}
    \end{minipage}
    \begin{minipage}[t]{0.45\textwidth}
    \centering
    \caption{Norms of $d$-vector components $|d^\alpha|$ in the case of AFM fluctuations.
    Bold types represent predominant components.
    The $d$-vector of the $B_{1u}$ state is aligned nearly parallel to the $a$-axis.}
    \label{table:AFM}
    \scalebox{1.}[1.]{
    \begin{tabular}{ccccc}
        \hline\hline
            & $|\psi|$ & $|d^a|$ & $|d^b|$ & $|d^c|$ \\ \hline
        $A_u$  & 0.00   &  0.04   & 0.06   & $\bm{0.99}$   \\
        ~$B_{1u}$~  &  ~0.39~  &  $\bm{0.92}$~  & ~0.05~   &  ~0.03~  \\
        $B_{2u}$  & 0.00 & 0.33   & 0.05   &  $\bm{0.94}$     \\
        $B_{3u}$  & 0.04   & 0.00   & $\bm{0.99}$   &  0.06  \\
        $A_g$  & $\bm{0.99}$   & 0.03  &  0.05  &  0.00  \\
        $B_{1g}$  & $\bm{0.99}$   & 0.10  &  0.10  &  0.00  \\
        $B_{2g}$  & $\bm{0.99}$   & 0.03  &  0.10  &  0.01  \\
        $B_{3g}$ & $\bm{0.99}$   & 0.11   &  0.04  & 0.00   \\ \hline\hline
    \end{tabular}}
    \end{minipage}
\end{table}

\begin{figure}[t]
    \centering
    \includegraphics[width=\linewidth]{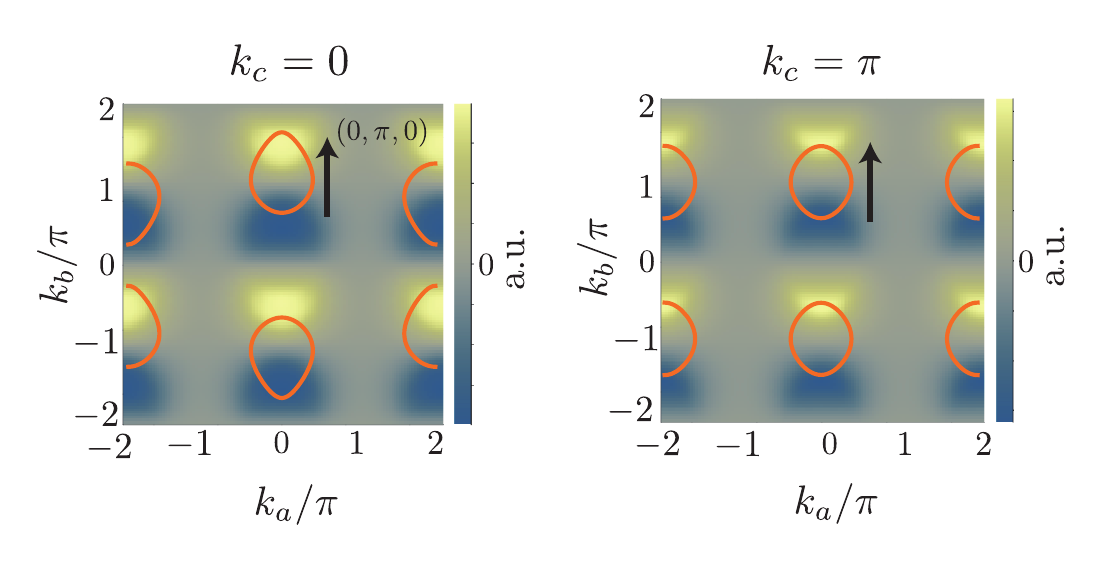}
    \caption{The $\bm{k}$-dependence of $d^a$ for the intraorbital pairings ($\tau_0$-component) of the $B_{1u}$ state which is induced by Ising AFM fluctuation. 
    The color bar represents $d^a_0(k_a,k_b,k_c,n=1)$ and the orange curves represent the electron Fermi surface.
    The sign of $d^a$ changes by shifting the momentum $\bm{k} \rightarrow \bm{k}+\bm{Q}$, where $\bm{Q}$ is illustrated by the bold arrow.
    The $d^a$ for the interorbital pairings ($\tau_x$-component) of the $B_{1u}$ state also exhibits similar $\bm{k}$-dependence. 
    }
    \label{fig:daB1u}
\end{figure}

In this section, we delve into the superconducting states induced by the interplay between
FM and  AFM fluctuations,
comparing the Eliashberg eigenvalues of all irreducible representations for pairing states.
First, we consider the case that the interaction between electrons is mediated solely by FM or AFM fluctuations.
The calculated results of  the largest eigenvalues of the Eliashberg equations are shown in Fig.~\ref{fig:FMAFM}. 
In Tables~\ref{table:FM} and \ref{table:AFM} , we show the numerical results of the norms of $d$-vector components, $|d^\alpha| = \sqrt{\sum_{\bm{k},n,j}|d^\alpha_j(\bm{k},i\omega_n)|^2}$ which are defined by the sum of all orbital contributions.\color{black}

In the case of FM fluctuations,
we set the interaction parameters as follows: $J^a = 30$,~$J^b=0$,~$J^c=0$, and $(1/{\xi^a})^2 = 0.1$.
The eigenvalues of the $A_u$ and $B_{3u}$ states are the largest among all the irreducible representations and are nearly degenerate as shown in Fig.~\ref{fig:FMAFM}(a).
For the $A_u$ ($B_{3u}$) state, the predominant $d$-vector component is $d^b \sim k_b$ ($d^c \sim k_b$) as shown in Table.~\ref{table:FM}.
Given the nearly degenerate $d$-vectors, they can rotate within the $bc$ plane.
These results are in agreement with a previous theoretical study~\cite{ishizuka2021Periodic}.
As is well known~\cite{monthoux1999Wave}, longitudinal FM fluctuations along the easy-axis ($a$-axis) stabilize the spin-triplet pairings with the magnetic moment along this axis.

In the case of AFM fluctuation, we set the interaction parameters as follows: $J^a = 100$,~$J^b=0$,~$J^c=0$, and $(1/{\xi^a})^2 = 0.1$.
As shown in Fig.~\ref{fig:FMAFM}(b) and Table~\ref{table:AFM}, the $B_{1u}$ state exhibits the largest eigenvalue, with the predominant $d$-vector component being $d^a\sim k_b$.
To elucidate why the $B_{1u}$ state is stabilized by AFM fluctuations,
we present heatmaps illustrating the $\bm{k}$-dependence of the $d^a$ component of the $B_{1u}$ state in Fig.~\ref{fig:daB1u}.
The sign of $d^a$ is changed by shifting the momentum $\bm{k} \rightarrow \bm{k}+\bm{Q}$~(illustrated in bold arrow).
Furthermore, according to Table~\ref{table:vertex}, the longitudinal fluctuations $\chi^a$ with a negative sign couple to $d^a$ in the Eliashberg equations.
Thus, the Ising AFM fluctuations give rise to an attractive interaction for the $d^a$ component.
This is analogous to the scenario of high-$T_c$ cuprate superconductors, wherein isotropic AFM fluctuations trigger spin-singlet superconductivity~\cite{monthoux1999Wave}.
However, in the case of the Ising AFM fluctuations without $\chi^b$ and $\chi^c$, the interacting channel is solely $\chi^a$.
Consequently, a one-dimensional ordering vector stabilizes the $p$-wave $B_{1u}$ pairing states more than even-parity pairing states.
In Appendix B, we will see that the even-parity pairing states are stabilized in the case with substantial transverse fluctuations, which is
in agreement with previous studies~\cite{ishizuka2021Periodic}.

The numerical results obtained above yield insights into the multiple superconducting phases of UTe$_2$ under applied pressure.
To address this issue, we consider the case with both FM and AFM fluctuations and examine stable pairing states, changing the correlation length of
these fluctuations.
We take into account the coexistence of FM and AFM fluctuations by assuming a simple form $\chi^a = \chi^a_{\text{FM}} + \chi^a_{\text{AFM}}$.
The origin of the AFM fluctuations observed in neutron scattering experiments remains unclear.
However, it is plausible to consider that they are deeply related to a magnetically ordered phase which appears under pressure higher than $1.5$GPa~\cite{aoki2020Multiple}. 
Hence, we postulate the presence of an antiferromagnetic critical point around $p_c = 1.5$GPa, and 
the correlation length of AFM fluctuations obeys the power law with the mean-field exponent,
\begin{eqnarray}
    \label{eq:power}
    \xi^a_{\text{AFM}}(p) = \frac{C}{(p_c - p)^{1/2}},
\end{eqnarray}
where $p$ is the pressure parameter, $p_c$ is the critical pressure, and a constant $C$ is set to $5$ for numerics.
On the other hand, to simplify the analysis, we hypothesize that weak FM fluctuations exist only close to ambient pressure, and
$\xi^a_{\text{FM}} = 4~(p<0.2)$ and  $\xi^a_{\text{FM}} = 1~(p>0.2)$.
The parameters $J^{\mu}$ are set as follows:
$J^a_{\text{FM}} = 10$,~$J^a_{\text{AFM}} = 100$,~$J^b_{\text{FM/AFM}}=0$, and $J^c_{\text{FM/AFM}}=0$.

\begin{figure}[t]
    \centering
    \includegraphics[width=\linewidth]{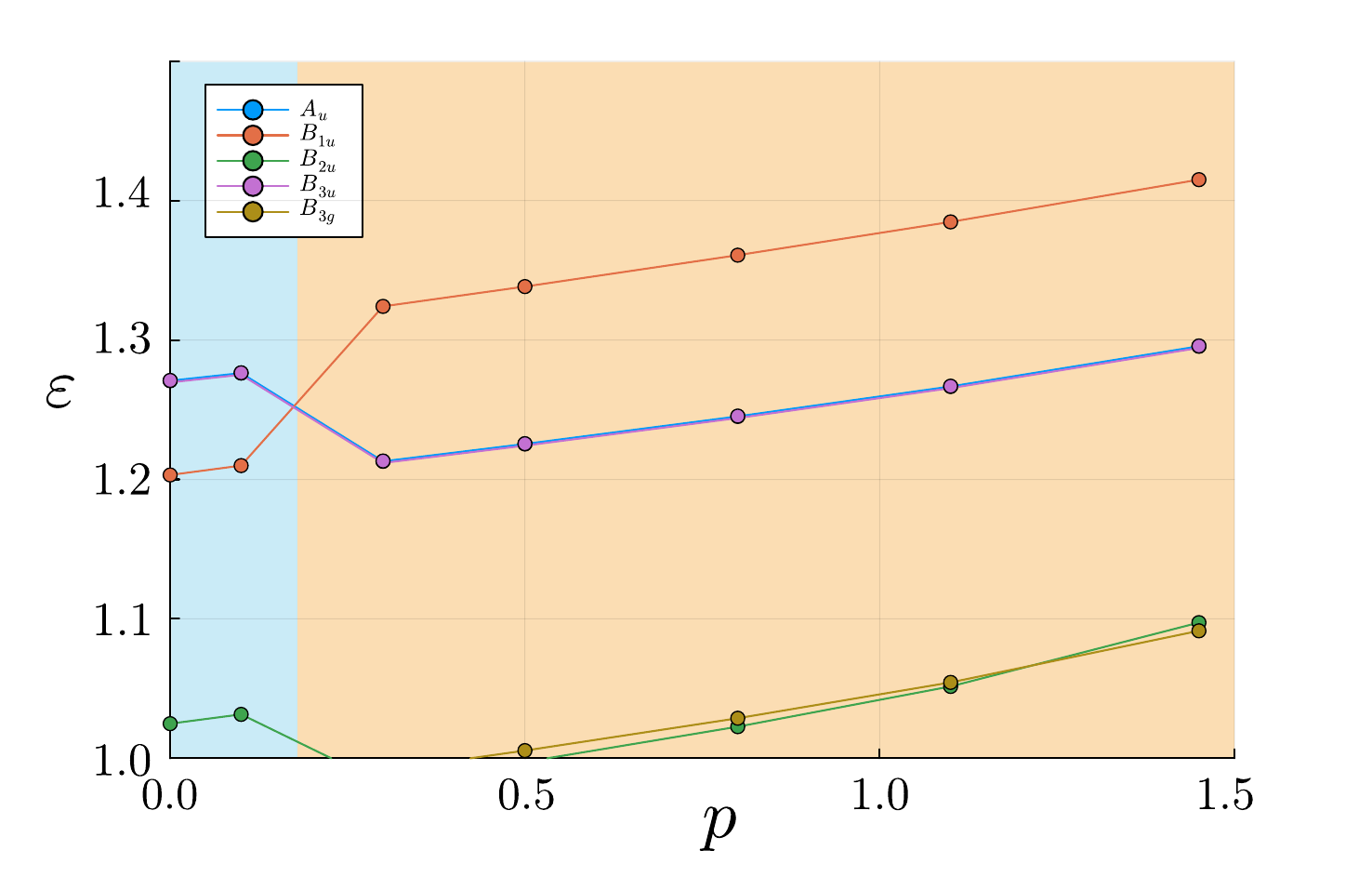}
    \caption{The largest eigenvalue of the Eliashberg equation for each irreducible representation versus the pressure parameter $p$.
    The correlation length of
    weak FM fluctuations is assumed to be $\xi^a_{\rm{FM}}=4$ for $p < 0.2$ and $\xi^a_{\text{FM}} = 1$ for $p>0.2$, 
   while the correlation length of AFM fluctuations obeys Eq.~(\ref{eq:power}). 
    The parameters $J^{\mu}$ are set as follows:
    $J^a_{\text{FM}} = 10$,~$J^a_{\text{AFM}} = 100$,~$J^b_{\text{FM/AFM}}=0$, and $J^c_{\text{FM/AFM}}=0$.
    At lower pressure $(p<0.2)$, the coexistence of FM and AFM fluctuations stabilize the nearly degenerate $A_u$ and $B_{3u}$ states,
    In contrast, at higher pressure $(p>0.2)$, AFM fluctuations stabilize the $B_{1u}$ state.
}
    \label{fig:scphase}
\end{figure}

Fig.~\ref{fig:scphase} illustrates the largest eigenvalue of the Eliashberg equation for each irreducible representation
as a function of the pressure parameter $p$.
At lower pressures, $p < 0.2$, the coexistence of FM and AFM fluctuations stabilizes the nearly degenerate $A_u$ and $B_{3u}$ states.
At higher pressures $p > 0.2$, since FM fluctuations are suppressed, whereas AFM fluctuations are enhanced,  the $B_{1u}$ state is stabilized.
These results provide a comprehensive understanding of the multiple superconducting phases of UTe$_2$ under pressure. 
Comparing the phase diagram shown in Fig.~\ref{fig:scphase}  and the experimental observation (Fig.~\ref{fig:model}(a)), we
can identify the SC1 phase with the $A_u$ or $B_{3u}$ state, and the SC2 phase with the $B_{1u}$ state.
Although we use the particular pressure-dependence of the correlation length, Eq.~(\ref{eq:power}), in the calculations, the qualitative behaviors of the phase diagram shown in Fig.~\ref{fig:scphase} are not changed by other choices of pressure-dependence of $\xi_{\rm FM/AFM}$, as long as the AFM fluctuations are weaker than the FM fluctuations at ambient pressure, and $\xi_{\rm AFM}$ ($\xi_{\rm FM}$) increases (decreases) substantially as the pressure increases.

\section{Summary and Discussion}
In this paper, we postulate that pairing glues are FM and AFM fluctuations,
and investigate the stability of superconducting states by analyzing the linearized Eliashberg equations.
Our findings reveal that for the two-orbital model with the cylindrical Fermi surface observed via the dHvA measurements,
Ising-like spin fluctuations give rise to spin-triplet superconductivity, irrespective of their origin from either FM or AFM characters.
Furthermore, our results provide a useful framework for understanding the multiple superconducting phases in UTe$_2$ based on the spin-fluctuation-mediated pairing mechanism.
According to the results shown in Fig.~\ref{fig:scphase}, the SC1 phase in Fig.~\ref{fig:model}(a) is either the $A_u$ or $B_{3u}$ states, while
the SC2 phase under higher pressure is identified with the $B_{1u}$ state for which the $d$-vector is aligned almost parallel to the $a$-axis.

The recent NMR measurements under applied pressure~\cite{kinjo2023Superconducting} show that the Knight shift does not decrease below the superconducting transition temperature of the SC2 phase for a magnetic field along the $b$-axis, which is consistent with
our finding that the $d^b$ component is negligibly small.

The result of a spin-triplet pairing state induced by Ising AFM fluctuations raises
an interesting question about to what extent this result is general or model-dependent. 
As a matter of fact, our calculations indicate that this is model-dependent, since the stability of the spin-triplet $B_{1u}$ state strongly relies on
the matching of the $\bm{Q}$ vector and the shape of the Fermi surface, as depicted in Fig.~\ref{fig:daB1u}.
That is, within our theoretical framework, the nesting with the ordering vector $\bm{Q}$ occurs within a cylindrical Fermi surface, which stabilizes
the $\bm{d}$-vector of the $B_{1u}$ state. 

In this study, we mainly consider the case of Ising spin fluctuations. However, as discussed in Appendix B, if transverse spin fluctuations as well as
longitudinal spin fluctuations develop, stable pairing states may be changed: i.e.
transverse spin fluctuations comparable to longitudinal fluctuations stabilize spin-singlet pairing states, suppressing spin-triplet pairing states.
Thus, it is quite important to clarify experimentally the character of spin fluctuations of UTe$_2$ under pressure.
According to the measurement of magnetic susceptibilities under pressure~\cite{li2021Magnetica}, there is still strong Ising-like anisotropy even under high pressure of $1.4$ GPa, where the SC2 phase appears. 
Although the anisotropy of the static susceptibilities is not directly related to the anisotropy of spin fluctuations,
the origin of magnetic anisotropy, spin-orbit interactions, should be common between the static and dynamical magnetic properties, and hence, 
it is highly expected that even under high-pressure region, where the SC2 phase appears, transverse spin fluctuations are suppressed compared to longitudinal fluctuations.
The elucidation of the nature of spin fluctuations under pressure is a quite important future issue.
Finally, we comment on the recently proposed pairing mechanism resulting from the interplay between the Hund coupling and the Kondo exchange interactions~\cite{hazra2023Triplet}, 
distinct from the spin fluctuation-mediated mechanism.
The pairing symmetries generated by these two mechanisms are generally different and may be in competition in some cases.
Exploring this competition is an interesting future issue.\color{black}

\section*{ACKNOWLEDGMENTS}
This work was supported by JST CREST Grant No. JPMJCR19T5, Japan, 
a Grant-in-Aid for Scientific Research on Innovative Areas ``Quantum Liquid Crystals" (Grant No.~JP22H04480) 
and JSPS KAKENHI (Grant No.~JP20K03860, No.~JP21H01039, and No.~JP22H01221).

\begin{appendix}
\section{Green's function}
Here, we present the single-particle Green's function for the normal Hamiltonian in Eq.~(\ref{eq:Green}).
With the basis of orbital and spin $\tau \otimes \sigma$, the normal Hamiltonian Eq.~(\ref{eq:Hn}) is written in the form of the $4\times 4$ matrix,
\begin{eqnarray}
     \scalebox{0.9}{$H_N(\bm{k}) = \begin{pmatrix}
        \epsilon_0(\bm{k}) - \mu + \bm{g}(\bm{k})\cdot\boldsymbol{\sigma} & f_x(\bm{k}) - if_y(\bm{k}) \\
        f_x(\bm{k}) + if_y(\bm{k}) & \epsilon_0(\bm{k}) - \mu -  \bm{g}(\bm{k})\cdot\boldsymbol{\sigma}
    \end{pmatrix}$}.
\end{eqnarray}
This matrix can be readily diagonalized as follows:
\begin{eqnarray}
    &&U^\dagger(\bm{k}) H_N(\bm{k}) U(\bm{k}) = \begin{pmatrix}
        E_+(\bm{k})\hat{1} & \\
        & E_-(\bm{k})\hat{1}
    \end{pmatrix}, \\
    &&E_\pm(\bm{k}) = \epsilon_0(\bm{k}) -\mu \pm \sqrt{|\bm{g}(\bm{k})|^2 + |f(\bm{k})|^2},
\end{eqnarray}
\begin{widetext}
    \noindent where $f(\bm{k})=f_x(\bm{k})-if_y(\bm{k})$, $\hat{1}$ is a $2\times 2$ unit matrix, the unitary operator $U(\bm{k}) $ is
    \begin{eqnarray}
        U(\bm{k}) =  
        \begin{pmatrix}
            \cos \frac{\theta}{2} \cos \frac{\varphi}{2} & \sin \frac{\theta}{2} \sin \frac{\varphi}{2} e^{-i\psi} &  \cos \frac{\theta}{2} \sin \frac{\varphi}{2}e^{-i\phi} & \sin \frac{\theta}{2} \cos \frac{\varphi}{2} e^{-i\psi}e^{-i\phi} \\
            \sin \frac{\theta}{2} \cos \frac{\varphi}{2} e^{i\psi} & - \cos \frac{\theta}{2} \sin \frac{\varphi}{2} & \sin \frac{\theta}{2} \sin \frac{\varphi}{2} e^{i\psi} e^{-i\phi} & -\cos \frac{\theta}{2} \cos \frac{\varphi}{2}e^{-i\phi} \\
            \cos \frac{\theta}{2} \sin \frac{\varphi}{2} e^{i\phi} & \sin \frac{\theta}{2} \cos \frac{\varphi}{2} e^{-i\psi}e^{i\phi} & -\cos \frac{\theta}{2} \cos \frac{\varphi}{2} & -\sin \frac{\theta}{2} \sin \frac{\varphi}{2} e^{-i\psi} \\
            \sin \frac{\theta}{2} \sin \frac{\varphi}{2} e^{i\psi}e^{i\phi} & - \cos \frac{\theta}{2} \cos \frac{\varphi}{2} e^{i\phi} & - \sin \frac{\theta}{2} \cos \frac{\varphi}{2} e^{\i\psi} & \cos \frac{\theta}{2} \sin \frac{\varphi}{2}
        \end{pmatrix},
    \end{eqnarray}
    and,
    \begin{eqnarray}
        &&\tan \phi = \frac{f_y}{f_x}, ~~~
        \cos \varphi = \frac{|\bm{g}|}{\sqrt{|\bm{g}|^2 + |f|^2}}, \nonumber \\
        &&\cos \theta = \frac{g_a}{|\bm{g}|}, ~~~
        \tan \psi = - \frac{g_b}{g_c}. \nonumber
    \end{eqnarray}
The single-particle Green's function for non-interacting electrons, 
      $  G_0(\bm{k},i\omega_n) = [i\omega_n - H_N(\bm{k})]^{-1}$,
 is  expressed in the diagonalized basis as,
    \begin{eqnarray}
        G_0(\bm{k},i\omega_n) = U(\bm{k}) \begin{pmatrix}
            G^+_0(\bm{k},i\omega_n)\hat{1} & \\
            & G^-_0(\bm{k},i\omega_n)\hat{1}
        \end{pmatrix} U^\dagger(\bm{k}),
    \end{eqnarray}
    where $G^\pm_0(\bm{k},i\omega_n) = {1}/(i\omega_n - E_\pm(\bm{k}))$.
    We assume that the Fermi level resides solely within the lower band denoted by $E_-$,
    while the upper band is far away from the Fermi level. The parameters adopted in the main text indeed satisfy this condition.
    Then, we can reasonably neglect the contribution of the upper band channel  in the Green's function. 
    Within this approximation, we arrive at,
    \begin{eqnarray}
        \label{eq:green}
        \scalebox{0.8}{$        
        G_0(\bm{k},i\omega_n)
        = G^{-}_0(\bm{k},i\omega_n)
        \begin{pmatrix}
            \cos^2\frac{\theta}{2}\sin^2\frac{\varphi}{2} + \sin^2\frac{\theta}{2}\cos^2\frac{\varphi}{2} & -\frac{1}{2}\sin\theta\cos\varphi e^{-i\psi} & -\frac{1}{2}\sin\varphi e^{-i\phi} & 0 \\
            -\frac{1}{2}\sin\theta\cos\varphi e^{i\psi} & \cos^2\frac{\theta}{2}\cos^2\frac{\varphi}{2} + \sin^2\frac{\theta}{2}\sin^2\frac{\varphi}{2} &  0 & -\frac{1}{2}\sin\varphi e^{-i\phi} \\
            -\frac{1}{2}\sin\varphi e^{i\phi} & 0 & \cos^2\frac{\theta}{2}\cos^2\frac{\varphi}{2} + \sin^2\frac{\theta}{2}\sin^2\frac{\varphi}{2} &  \frac{1}{2}\sin\theta\cos\varphi e^{-i\psi} \\
            0 & -\frac{1}{2}\sin\varphi e^{i\phi} & \frac{1}{2}\sin\theta\cos\varphi e^{i\psi} & \cos^2\frac{\theta}{2}\sin^2\frac{\varphi}{2} + \sin^2\frac{\theta}{2}\cos^2\frac{\varphi}{2}
        \end{pmatrix}$}. \nonumber \\
    \end{eqnarray} 

    The full single-particle Green's function used for the calculations of superconducting states is defined as,
$G(\bm{k},i\omega_n)=[G_0(\bm{k},i\omega_n)^{-1}-\Sigma (\bm{k},i\omega_n) ]^{-1}$, where
$\Sigma (\bm{k},i\omega_n) $ is the single-particle self-energy of $f$-electrons expressed as Eq.~(\ref{eq:self}).
\end{widetext}

\begin{figure}[t]
    \begin{minipage}[t]{0.48\textwidth}
    \begin{center}
    \includegraphics[width=\linewidth]{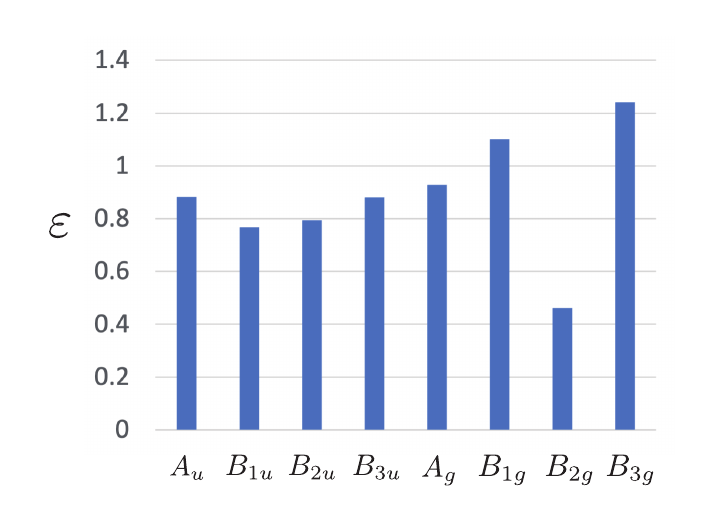}
    \caption{The largest eigenvalue of the Eliashberg equation of each irreducible representation in the case with both longitudinal and transverse AFM fluctuations.
         The interaction parameters are set as follows: $J^a_{\text{AFM}} = 100$,~$J^b_{\text{AFM}} = 30$,~$J^c_{\text{AFM}} = 30$,~$(1/{\xi^{a}})^2 = 0.1$,~$(1/{\xi^{b}})^2 = 0.1$,~$(1/{\xi^{c}})^2 = 0.1$, and $T_{\rm{sf}} = 4$.
         In the presence of transverse AFM fluctuations, the $B_{1u}$ state becomes unstable, whereas even-parity irreducible representations $B_{3g}$ and $B_{1g}$ become much more stable.}
    \label{fig:AFM2}
    \end{center}
    \end{minipage}

    \begin{minipage}{0.5\textwidth}
    \begin{center}
    \makeatletter
    \def\@captype{table}
    \makeatother
    \caption{Norms of $d$-vector components $|d^\alpha|$ in the case with both longitudinal and transverse AFM fluctuations.
    Bold types represent predominant components.
    The $d$-vector of the $B_{1u}$ state is not aligned along the $a$-axis in contrast to the case with longitudinal fluctuations only.}
    \begin{tabular}{ccccc}
        \hline\hline
            & $|\psi|$ & $|d^a|$ & $|d^b|$ & $|d^c|$ \\ \hline
        ~$B_{1u}$~  &  ~{0.57}~  &  {0.04}~  & ~$\bm{0.82}$~   &  ~0.05~  \\
        $B_{1g}$  & $\bm{0.99}$   & 0.03  &  0.06  &  0.00  \\
        $B_{3g}$ & $\bm{0.99}$   & 0.03   &  0.03  & 0.00   \\ \hline\hline
    \end{tabular}
    \label{table:AFM2}
    \end{center}
    \end{minipage}
\end{figure}
\section{Effects of transverse fluctuations}

\begin{figure}[t]
    \centering
    \includegraphics[width=\linewidth]{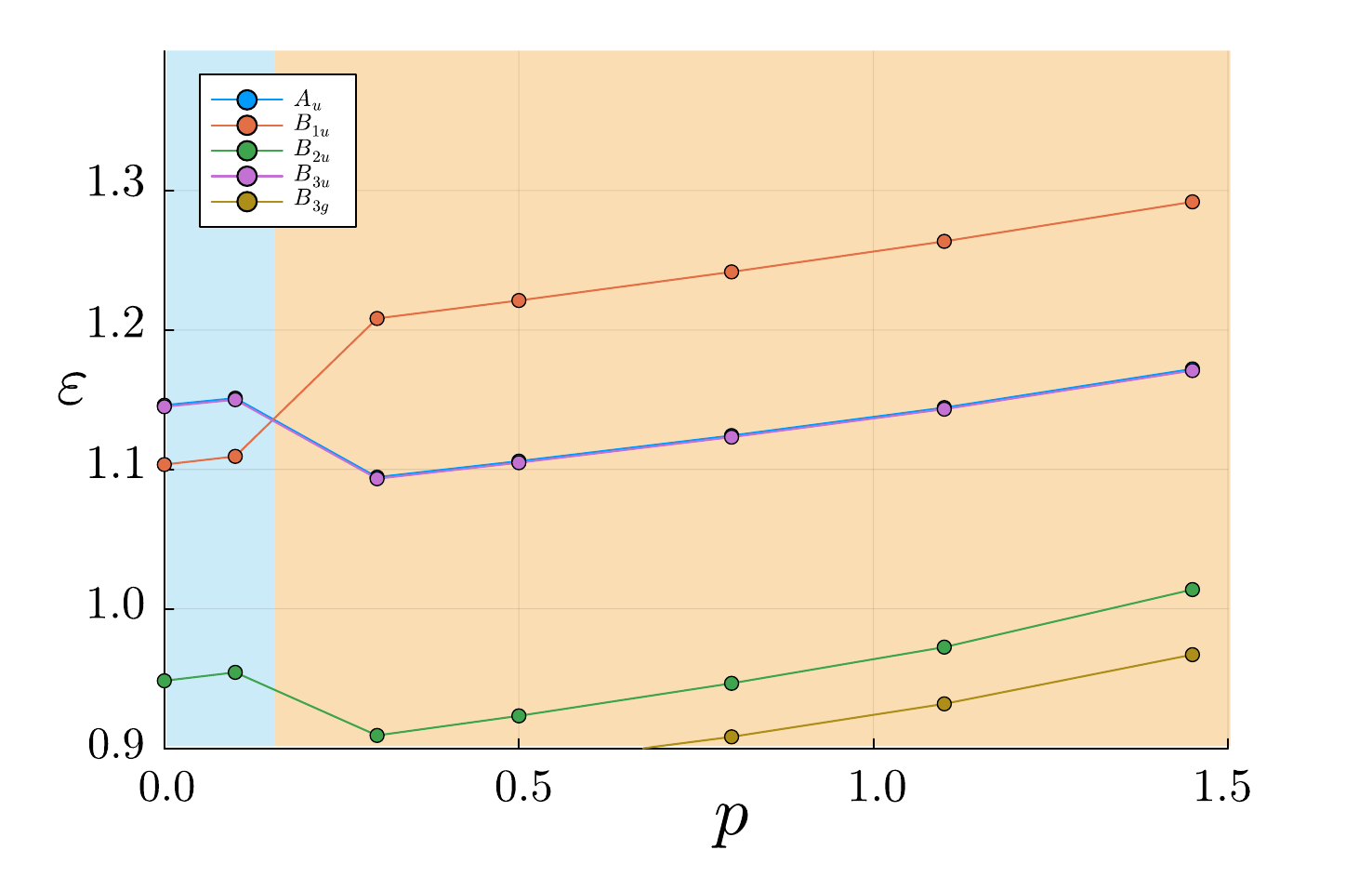}
    \caption{The largest eigenvalue of the Eliashberg equation of each irreducible representation versus the pressure parameter $p$ for scenario I.
    In addition to the longitudinal fluctuations considered in the main text,
    we assume the presence of weak transverse fluctuations with $\xi^b_{\text{FM/AFM}} = 2$ and $\xi^c_{\text{FM/AFM}} = 2$.
    The parameters $J^{\mu}$ are set as follows:
$J^a_{\text{FM}} = 10$,~$J^a_{\text{AFM}} = 100$,~$J^b_{\text{FM/AFM}}=10$, and $J^c_{\rm{FM/AFM}}=10$.
    The results are qualitatively similar to those in the case with only longitudinal fluctuations shown in Fig.~\ref{fig:scphase}.}
    \label{fig:scphase2}
\end{figure}
\begin{figure}[t]
    \centering
    \includegraphics[width=\linewidth]{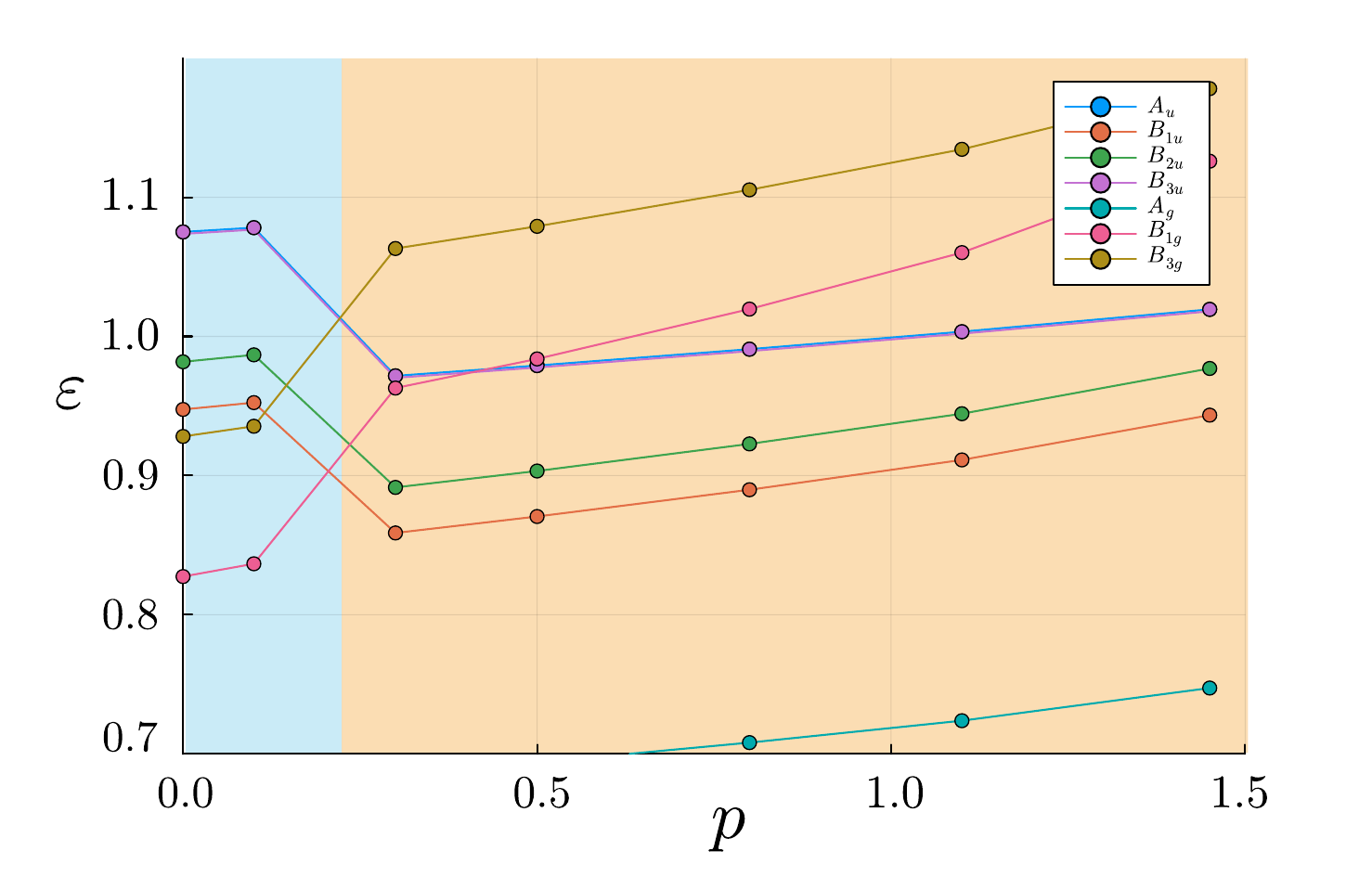}
    \caption{The largest eigenvalue of the Eliashberg equation of each irreducible representation versus the pressure parameter $p$ for scenario II.
   In addition to the longitudinal fluctuations considered in the main text,
    strongly developed transverse AFM fluctuations with the correlation length given by Eq.~(\ref{eq:power2}) are included.
    The parameters are set as follows:
    ~$J^a_{\rm{FM}}=20$,~$J^a_{\rm{AFM}}=100$,~$J^b_{\rm{FM}}=10$,~$J^b_{\rm{AFM}}=30$,~$J^c_{\rm{FM}}=10$, and $J^c_{\rm{AFM}}=30$.
    At higher pressure $(p>0.2)$, the even-parity $B_{3g}$ state is most stable.
}
    \label{fig:scphase3}
\end{figure}

Although we consider the case with strong Ising anisotropy of spin fluctuations in the main text, 
it is important to examine the effects of transverse spin fluctuations, which may exist to some extent in UTe$_2$. 
Particularly, our computations reveal that Ising AFM fluctuations induce spin-triplet pairings,
whereas conventional wisdom posits that isotropic AFM fluctuations stabilize spin-singlet pairings.
In fact, a previous study for UTe$_2$~\cite{ishizuka2021Periodic} has reported that AFM fluctuations induce an even-parity irreducible representation
when the anisotropy of spin fluctuations becomes weaker.

We consider the case of anisotropic AFM fluctuations with weak transverse fluctuations.
We set the interaction parameters as follows:
$J^a_{\rm{AFM}} = 100$,~$J^b_{\rm{AFM}}=30$,~$J^c_{\rm{AFM}}=30$, and $(1/{\xi^\mu_{\rm{AFM}}})^2 = 0.1~(\mu=a,b,c)$.
We show the largest eigenvalues of all the irreducible representations in Fig.~\ref{fig:AFM2},
and also the norms of $d$-vector components of $B_{1u}$, $B_{1g}$, and $B_{3g}$ in Table~\ref{table:AFM2}.
In these calculations, the even-parity state $B_{3g}$ exhibits the largest eigenvalue among all the  irreducible representations.
Notably, the $B_{1u}$ state is not stabilized by AFM fluctuations, and its $d$-vector is not aligned along the $a$-axis as shown in Table~\ref{table:AFM2}, in contrast to the case of Ising fluctuations in the main text.
This is because the longitudinal fluctuations $\chi^a$, and the transverse fluctuations $\chi^b$ and $\chi^c$,
cancel each other for the channel of the $d^a$ component as shown in Table~\ref{table:vertex}.
On the other hand, all fluctuations, $\chi^a$, $\chi^b$, and $\chi^c$, act to stabilize the spin-singlet components $\psi$.
Consequently, AFM fluctuations with a certain degree of transverse fluctuations stabilize spin-singlet pairings.

Based on this result, we examine two scenarios of the multiple superconducting phases under pressure.
In the first scenario (scenario I), which is a slight modification of the scenario in the main text, we assume the existence of AFM longitudinal fluctuation along the $a$-axis with the correlation length given by Eq.~(\ref{eq:power})
and the existence of weak ferromagnetic fluctuations along the $a$-axis characterized by $\xi^a_{\text{FM}}=4~(p<0.2)$ and $\xi^a_{\text{FM}}=1~(p>0.2)$.
Additionally, we assume the presence of weak transverse fluctuations characterized by $\xi^b_{\text{FM/AFM}} = 2$ and $\xi^c_{\text{FM/AFM}} = 2$.
The parameters $J^{\mu}$ are set as follows:
$J^a_{\text{FM}} = 10$,~$J^a_{\text{AFM}} = 100$,~$J^b_{\text{FM/AFM}}=10$, and $J^c_{\rm{FM/AFM}}=10$.
In Fig.~\ref{fig:scphase2}, we show the largest eigenvalues of each irreducible representation as a function of pressure.
The result remains qualitatively similar to the case without transverse fluctuations considered in the main text.
At lower pressure ($p<0.2$), FM fluctuations stabilize the nearly degenerate $A_u$ and $B_{3u}$ states,
while at higher pressure ($p>0.2$), Ising AFM fluctuations stabilize the $B_{1u}$ state.

On the other hand, in the second scenario (scenario II), in addition to the strong AFM longitudinal fluctuation,
we consider the comparable transverse AFM fluctuations as well, with their correlation length characterized by the power law,
\begin{eqnarray}
    \label{eq:power2}
    \xi^{\mu}_{\rm{AFM}}(p) = \frac{C}{(p_c -p)^{1/2}} ~~~ \mu = a,b,c.
\end{eqnarray}
We also assume the existence of weak ferromagnetic fluctuations with $\xi^a_{\text{FM}}=4~(p<0.2)$, $\xi^a_{\text{FM}}=1~(p>0.2)$,
$\xi^b_{\text{FM}} = 2$ and $\xi^c_{\text{FM}} = 2$.
The parameters $J^{\mu}$ are set as follows:
$J^a_{\text{FM}} = 20$,~$J^a_{\text{AFM}} = 100$,~$J^b_{\text{FM}}=10$,~$J^b_{\text{AFM}}=30$,~$J^c_{\text{FM}}=10$, and $J^c_{\text{AFM}}=30$.
Note that these spin fluctuations are still anisotropic $\chi^a > \chi^b, \chi^c$.
Fig.~\ref{fig:scphase3} illustrates the largest eigenvalues of all the irreducible representations as a function of pressure.
At lower pressure ($p < 0.2$), FM fluctuations stabilize the nearly degenerate $A_u$ and $B_{3u}$ states.
In contrast, at higher pressure ($p > 0.2$),  where transverse fluctuations develop, the $B_{3g}$ state is the most stable.

\end{appendix}

\bibliography{paper}

\end{document}